\documentclass[preprint,pra]{revtex4-2}
\usepackage{makeidx}
\usepackage{amssymb}
\usepackage{amsmath}
\usepackage{eurosym}
\usepackage{graphicx}
\usepackage{dcolumn}
\usepackage{bm}

\begin{document}

\title{Solitons in optical couplers: introduction and perspectives}
\author{Boris A. Malomed}
\address{Department of Physical Electronics, School of Electrical and Computer
Engineering, Faculty of Engineering, Tel Aviv University, Tel Aviv
69978, Israel}

\begin{abstract}
This minireview provides a brief summary and a discussion of directions for
further development of theoretical and, chiefly, experimental studies of
bright solitons in optical couplers, i.e., dual-core waveguides which
combine the linear inter-core coupling (tunneling of light between the
parallel cores with the intra-core group-velocity dispersion and
self-focusing Kerr (cubic) nonlinearity.\ Following a short introduction to
the field, the article focuses on a brief review of relatively recent
experimental results for the switching of solitons in dual-core nonlinear
optical fibers and the spontaneous emergence of stable \textit{asymmetric}
two-core solitons in the couplers with the \textit{symmetric} dual-core
structure.
\end{abstract}

\maketitle

\section{Introduction}

\begin{figure}[tbp]
\begin{center}
\includegraphics[height=6cm]{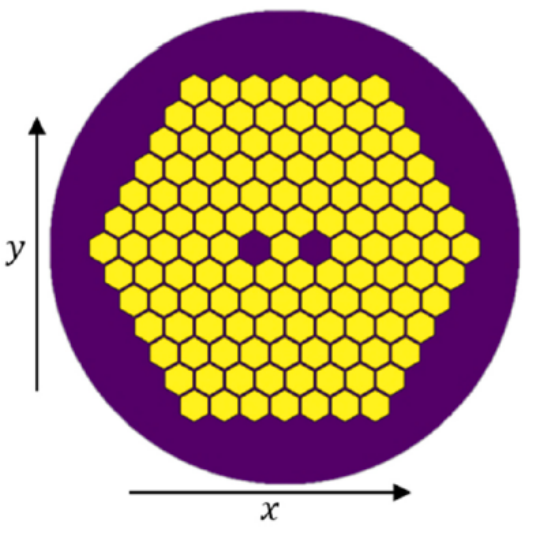}
\end{center}
\caption{A schematic transverse profile of a microstructured optical fiber,
with two tunnel-coupled waveguding cores created in it, as per Ref.
\protect\cite{Longo}.}
\label{structured}
\end{figure}

In the great variety of optical waveguides, one of basic types represents
dual-core \textit{couplers}, in which propagating optical waves tunnel
between parallel guiding cores via evanescent fields tunneling across the
dielectric barrier separating the cores \cite{Huang}. In most cases, the
couplers are realized as twin-core optical fibers \cite%
{fiber-coupler,fiber-coupler2,Old}. In a more sophisticated form, these may
be inner twin-core structures in photonic-crystal fibers \cite{PCF}. The
dual-core fibers are fabricated by drawing an appropriately shaped preform
from melt, or pressing together two single-mode fibers. Technology for the
fabrication of a microstructured fiber with an inner dual guiding core is
available too \cite{Russell}, see Fig. \ref{structured}.

The wave exchange between the parallel cores in the coupler is affected by
the intra-core nonlinearity \cite{Jensen}. This effect has been employed to
design diverse all-optical switching devices \cite{switch1}-\cite{Ignac3}
and other applications, such as nonlinear amplifiers \cite{amplifier}, logic
gates \cite{logic}, and bistability \cite{Leon}. Nonlinear couplers can also
provide efficient compression of solitons by transferring it into a fiber
with a smaller value of the group-velocity dispersion (GVD) coefficient: the
highest compression quality is achieved when two fibers with different GVD
coefficients are not directly spliced one into the other, but are, instead,
parallel-connected in the form of a coupler \cite{HH}.

In addition to the basic dual-core system, the implementation of nonlinear
couplers has been proposed in many other photonic settings, including light
with the polarization structure \cite{Trillo}, semiconductor waveguides \cite%
{semi}, plasmonic media \cite{plasma1,plasma2,plasma3}, twin-core fiber
Bragg gratings \cite{Bragg,Sukhorukov}, and others. In addition to the
ubiquitous Kerr (cubic) nonlinearity of the core material, the analysis has
been developed for systems with nonlinearities of other types, represented
by saturable \cite{satur}, quadratic (alias second-harmonic-generating) \cite%
{chi2}, cubic-quintic (CQ) \cite{Albuch}, and nonlocal cubic terms \cite%
{Nonlocal}. Double-core transmission of light can be realized not only in
optical fibers, but also in dual-core planar waveguides (i.e., in the
\textit{spatial domain}, rather than the \textit{temporal domain}) \cite%
{Smirnova}. In most cases, the nonlinear dual-core waveguides are adequately
modeled by systems of linearly coupled nonlinear Schr\"{o}dinger equations
(NLSEs), in which the tunneling of light between the cores is accounted for
by linear-coupling terms \cite{Jensen,Wabnitz,Snyder}. The coupler concept
was also extended for the \textit{spatiotemporal }propagation of light in
dual-core planar waveguides, with the respective one-dimensional NLSE system
replaced by its two-dimensional version, which includes the temporal and
spatial coordinates \cite{Dror}.

A fundamental property of nonlinear couplers with mutually symmetric twin
cores is the \textit{symmetry-breaking bifurcation} (SBB), which
destabilizes obvious symmetric modes and gives rise to asymmetric ones. The
SBB was first analyzed for uniform states (alias \textit{continuous waves},
CWs) in dual-core nonlinear optical fibers \cite{Snyder}, and then for
solitons in the same system \cite{Wabnitz}-\cite{Skinner}, as well as for
dual-core nonlinear fibers with Bragg gratings (BGs) written on each core
\cite{Bragg}. Some theoretical results on this topic were summarized in an
early review \cite{Wabnitz2}, and later in Ref. \cite{Progress}. The SBB\
analysis was then extended to solitons in couplers with the quadratic \cite%
{chi2} and CQ \cite{Albuch} nonlinearities.

The cubic (Kerr) nonlinearity in the dual-core system gives rise to the
\emph{subcritical} SBB for solitons. This means that it originally produces
unstable branches of asymmetric modes which evolve \emph{backward} in the
parameter plane of the soliton's energy and asymmetry degree (in the
direction of smaller energy. i.e., weaker nonlinearity), see Fig. \ref{fig2}
below. Then, the branches turn forward, retrieving stability at the turning
points (the bold ones in Fig. \ref{fig2}) \cite{bifurcations}. On the other
hand, the \emph{supercritical} SBB immediately gives rise to stable branches
of asymmetric solitons, which evolve strictly in the forward direction. For
solitons in couplers, the SBB of the latter type\ occurs in twin-core BGs
\cite{Bragg}, and in the system with the quadratic nonlinearity \cite{chi2}.

In\ the model of the dual-core coupler with the cubic intra-core
nonlinearity, the SBB point for symmetric solitons was found in an exact
analytical form \cite{Wabnitz}.The asymmetric solitons emerging at the SBB\
point were studied analytically by means of the variational approximation
(VA) \cite{Laval,satur,chi2,Bragg,Dror} and in the numerical form \cite%
{Akhm,Akhm2} (in addition to the original works which reported these
findings, see also a summary in reviews \cite{Wabnitz2}, \cite{Progress},
and \cite{Old}).

While the theoretical analysis of solitons in dual-core optical systems was
elaborated in detail, starting in 1989 \cite{Wabnitz}-\cite{Akhm2}, the
first experimental realization of this phenomenology in optical fibers was
reported only in 2020 \cite{Ignac1}, which initiated renewed interest to
this topic and stimulated further experimental studies \cite%
{Ignac,Ignac2,Ignac3}. The objective of the present mini-review is to
briefly summarize the experimental findings for the solitons, as a topic
with considerable potential for further work. Because these findings are
relatively novel ones, they were not included in the previous reviews on the
theme of solitons in dual-core systems \cite{Wabnitz2,Progress,Old}.

The subsequent presentation is organized as follows. First, a necessary
summary of well-known theoretical results, which are focused on the
prediction of the SBB in nonlinear fiber-optic couplers, is presented in a
brief form in Section 3. This is followed by the core part of the article,
\textit{viz}., the summary of the relatively recent experimental findings
for the solitons. The article is concluded by Section 4.

\section{Solitons in dual-core nonlinear fibers: the basic
(well-established) theory}

The basic model of the symmetric coupler amounts to the system of linearly
coupled NLSEs \cite{fiber-coupler2}:
\begin{eqnarray}
iu_{z}+\frac{1}{2}u_{\tau \tau }+|u|^{2}u+Kv &=&0,  \label{ucoupler} \\
iv_{z}+\frac{1}{2}v_{\tau \tau }+|v|^{2}v+Ku &=&0.  \label{vcoupler}
\end{eqnarray}%
Here $z$ is the propagation distance, $\tau \equiv t-z/V_{\mathrm{gr}}$ is
the reduced time ($t$ is the physical time, and $V_{\mathrm{gr}}$ is the
group velocity of the carrier wave \cite{Agrawal}), $u$ and $v$ are
envelopes of the optical waves in the coupled cores, the second derivatives
represent the anomalous GVD \cite{Agrawal}, the cubic terms represent the
Kerr nonlinearity in the cores, and real $K$ is the coupling constant
accounting for the tunneling of light between the cores, which may always be
defined to be positive.\

Equations (\ref{ucoupler}) and (\ref{vcoupler}) can be derived from the
respective Lagrangian ($L$), which is a combination of the kinetic terms and
the Hamiltonian ($H$) \cite{Progress}:
\begin{eqnarray}
L &=&\int_{-\infty }^{+\infty }\left[ \frac{i}{2}\left( u^{\ast
}u_{z}+v^{\ast }v_{z}\right) d\tau +\mathrm{c.c.}\right] -H,  \label{L} \\
H &=&\int_{-\infty }^{+\infty }\left[ \frac{1}{2}\left( \left\vert u_{\tau
}\right\vert ^{2}+\left\vert v_{\tau }\right\vert ^{2}\right) -\frac{1}{2}%
\left( |u|^{4}+|v|^{4}\right) -K\left( u^{\ast }v+uv^{\ast }\right) \right] ,
\label{H}
\end{eqnarray}%
where both the asterisk and c.c. stand for the complex conjugate. The
Hamiltonian, along with the energy (alias the total norm) of the solution,%
\begin{equation}
E=\frac{1}{2}\int_{-\infty }^{+\infty }\left( |u|^{2}+|v|^{2}\right) d\tau
\equiv E_{u}+E_{v},  \label{E}
\end{equation}%
and the total momentum,%
\begin{equation}
P=\frac{i}{2}\int_{-\infty }^{+\infty }\left[ \left( uu_{\tau }^{\ast
}+vv_{\tau }^{\ast }\right) +\mathrm{c.c.}\right] d\tau ,  \label{P}
\end{equation}%
are dynamical invariants of the system. Note that the energy is different
from the Hamiltonian, in the present context.

Equations (\ref{ucoupler}) and (\ref{vcoupler}) admit obvious symmetric and
antisymmetric soliton solutions,%
\begin{equation}
u=\pm v=T^{-1}\mathrm{sech}\left( \frac{\tau }{T}\right) \exp \left( \frac{iz%
}{2T^{2}}\pm iKz\right) ,  \label{+-}
\end{equation}%
where $T$ is the temporal width of the soliton, which determines its energy
and Hamiltonian:%
\begin{equation}
E_{\pm }=2T^{-1},~H_{\pm }=-\frac{2}{3}T^{-3}\mp 4KT^{-1}.  \label{Esymm}
\end{equation}%
The antisymmetric solitons make the coupling term in Hamiltonian (\ref{Esymm}%
) positive, unlike the negative one for the symmetric states. For this
reason, the antisymmetric solitons always correspond to a maximum, rather
than minimum, of the Hamiltonian, which makes them always unstable \cite%
{Akhm2}, Therefore, the antisymmetric modes are not considered below

The issue of major interest is the SBB, which destabilizes the symmetric
solitons and creates stable asymmetric ones, with unequal energies of their
components, but still identical signs of the corresponding wave fields,
inherited from the symmetric states. The value of the energy of the
symmetric soliton at the SBB point is known in an exact form \cite{Wabnitz}:%
\begin{equation}
E=E_{\mathrm{bif}}\equiv 4\sqrt{K/3}\approx 2.31\sqrt{K}.  \label{E2}
\end{equation}

Asymmetric solitons, which emerge at the SBB point, cannot be found in an
exact analytical form, but they can be studied by means of VA \cite{Laval}-%
\cite{Skinner}.This approximation is based on the following trial analytical
form (\emph{ansatz}) for the two-component soliton solution of Eqs. (\ref%
{ucoupler}) and (\ref{vcoupler}):
\begin{eqnarray}
u &=&A\cos (\theta )\mathrm{sech}\left( \frac{\tau }{T}\right) \exp \left(
i(\phi +\psi )+ib\tau ^{2}\right) ,  \label{uansatzcoupler} \\
v &=&A\sin (\theta )\mathrm{sech}\left( \frac{\tau }{T}\right) \exp \left(
i(\phi -\psi )+ib\tau ^{2}\right) ,  \label{vansatzcoupler}
\end{eqnarray}%
where the real shape parameters $A$ and $T$ are the overall amplitude and
temporal width of the soliton, while real $\theta $, provided that it takes
values $\neq \pi /4$, determines the asymmetry between the components, in
terms of the relative difference of their energies (see Eq. (\ref{E})),%
\begin{equation}
\cos (2\theta )\equiv \frac{E_{u}-E_{v}}{E_{u}+E_{v}}.
\label{couplerasymmetry}
\end{equation}%
Further, $\phi (z)$ and $\psi (z)$ are, respectively, overall and relative
phases of the two-component soliton, while real \emph{chirp} $b(z)$ must be
introduced if the ansatz admits variation of the width as a function of $z$
\cite{Anderson,Anderson2}. The asymmetric solitons with $E_{u}>E_{v}$ and $%
E_{u}<E_{v}$, i.e., $\theta >\pi /4$ and $\theta <\pi /4$, respectively (see
Eq. (\ref{couplerasymmetry})), may be considered as the asymmetric modes
with \emph{opposite polarities}.

Switching of a soliton between the two cores of the coupler was considered,
in the framework of a full system of variational equations for ansatz (\ref%
{uansatzcoupler},\ref{vansatzcoupler}) in Ref. \cite{Uzunov}. For static
solitons, ansatz (\ref{uansatzcoupler},\ref{vansatzcoupler}) with $b=0$
gives rise to the following variational equations, in which all parameters
of ansatz (\ref{uansatzcoupler},\ref{vansatzcoupler}), except for the phase $%
\phi (z)$, are assumed constant:
\begin{gather}
\sin (2\theta )\sin (2\psi )=0,  \label{coupler1} \\
\frac{E}{3T}\cos (2\theta )-K\,\mathrm{cot}(2\theta )\cos (2\psi )=0,
\label{coupler2} \\
T^{-1}=E\left[ 1-\frac{1}{2}\sin ^{2}(2\theta )\right] ,  \label{coupler3}
\end{gather}%
\begin{equation*}
\frac{d\phi }{dz}\,=\,-\,\frac{1}{6T^{2}}+\frac{2E}{3T}\left( 1-\frac{1}{2}%
\sin ^{2}(2\theta )\right) +\kappa \sin (2\theta )\cos (2\psi ).
\end{equation*}%
Here $E$ is the soliton's total energy, which, according to its definition (%
\ref{E}), takes value $E=A^{2}T$ for ansatz (\ref{uansatzcoupler},\ref%
{vansatzcoupler}).

As it follows from Eq. (\ref{coupler1}), the static asymmetric soliton,
i.e., one with $\theta \neq \pi /4$ (see Eq. (\ref{couplerasymmetry})), has $%
\sin (2\psi )=0$, which implies that $\cos (2\psi )=\pm 1$. As mentioned
above, the solutions corresponding to $\cos (2\psi )=-1$, i.e.,
antisymmetric ones, with respect to the two components, are unstable.
Therefore, only the case of $\cos (2\psi )=+1$, corresponding to solitons
with in-phase components ($\psi =0$), is relevant. Then, the soliton's width
$T$ can be eliminated by means of Eq. (\ref{coupler3}), and the remaining
equation (\ref{coupler3}) for the energy-distribution angle $\theta $ takes
the form of
\begin{equation}
\cos (2\theta )\left[ \frac{E^{2}}{3K}\sin (2\theta )\left( 1-\frac{1}{2}%
\sin ^{2}(2\theta )\right) -1\right] =0.  \label{couplertheta}
\end{equation}

Further analysis reveals that, in the interval $0<E<E_{1}$, where
\begin{equation}
E_{1}=\sqrt{(9/4)\sqrt{6}K}\approx \allowbreak 2.\,\allowbreak 348\,\sqrt{K},
\label{couplerE1}
\end{equation}%
the only relevant solution to Eq. (\ref{couplertheta}) is the symmetric one,
with $\theta =\pi /4$ and equal energies in both components, according to
Eqs. (\ref{uansatzcoupler},\ref{vansatzcoupler}) and (\ref{couplerasymmetry}%
). When the soliton's energy attains value $E_{1}$, predicted by Eq. (\ref%
{couplerE1}), \emph{asymmetric} solutions emerge by a jump from $\cos \left(
2\theta \right) =0$ to with
\begin{equation}
\cos (2\theta )=\pm 1/\sqrt{3}.  \label{asymmearliest}
\end{equation}%
When $E$ attains a slightly larger value,
\begin{equation}
E_{2}=\sqrt{6K}\approx 2.450\sqrt{K},  \label{couplerE2}
\end{equation}%
the above-mentioned \emph{backward (subcritical) bifurcation} \cite%
{bifurcations} occurs, which makes the symmetric solution with $\theta =\pi
/4$ unstable. The (slightly) subcritical character of the bifurcation, with
respect to the variation of energy $E$, is confirmed by the fact that the
asymmetric soliton (\ref{asymmearliest}) appears at the energy value (\ref%
{couplerE1}) which is \emph{slightly smaller} than the value of $E$ at the
bifurcation point (\ref{couplerE2}). The comparison with the full numerical
results corroborates the weakly subcritical shape of the SBB for solitons in
the nonlinear coupler \cite{Skinner}.

\begin{figure}[tbp]
\begin{center}
\includegraphics[height=6cm]{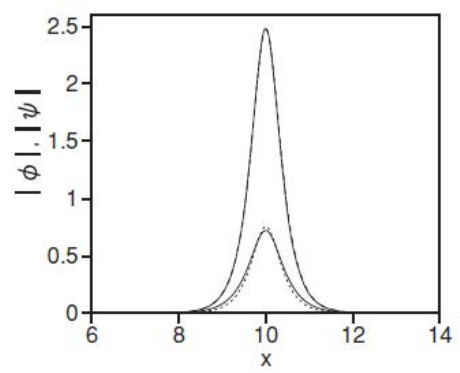}
\end{center}
\caption{A typical example of two components of a stable asymmetric soliton,
with $\left\vert \protect\phi (x)\right\vert \equiv U(\protect\tau )$, $%
\left\vert \protect\psi (x)\right\vert \equiv V(\protect\tau )$, as per Ref.
\cite{HS}. Continuous and dashed lines designate the numerically
found solution and its VA-produced counterpart, respectively.}
\label{fig1}
\end{figure}

\begin{figure}[tbp]
\begin{center}
\includegraphics[height=6cm]{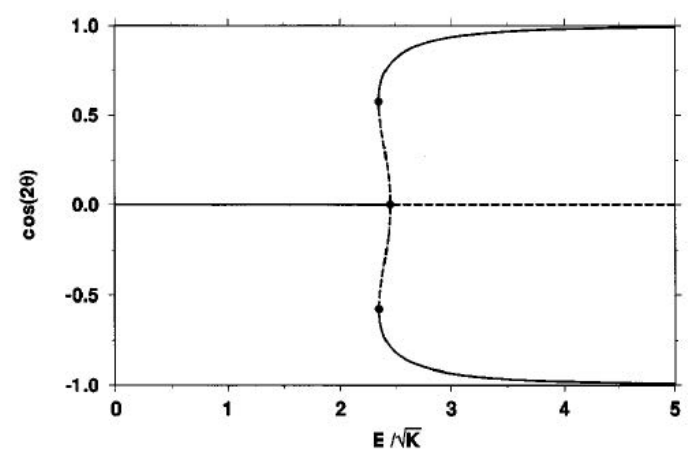}
\end{center}
\caption{The dependence of the asymmetry parameter $\cos (2\protect\theta )$
of the two-component solitons (see Eq. (\protect\ref{couplerasymmetry})), in
the nonlinear coupler with identical cores, on the scaled total energy, $E/%
\protect\sqrt{K}$, as predicted by the VA, see Eq. (\protect\ref%
{couplerasymmetry}). The figure demonstrates a weakly subcritical SBB, solid
and dashed lines designating stable and unstable states, respectively. The
results are presented as per Ref. \protect\cite{Skinner}.}
\label{fig2}
\end{figure}

A typical example of the asymmetric soliton is displayed in Fig. \ref{fig1},
and the SBB diagram is presented in Fig. \ref{fig2}. One can easily
distinguish between stable and unstable branches in the diagram, using
elementary theorems of the bifurcation theory \cite{bifurcations}. Note also
that the emergence of the pairs of stable and unstable asymmetric solitons
\textquotedblleft from nothing" at points (\ref{couplerE1}), which are
denoted by the bold dots in Fig. \ref{fig2}, is an elementary bifurcation of
another type, known as the \emph{saddle-node} bifurcation \cite{bifurcations}%
.

Thus, the VA predicts that the backward (subcritical) bifurcation occurs at
the value of the soliton's energy given by Eq. (\ref{couplerE2}). The
accuracy of the VA is characterized by comparison of this prediction with
the above-mentioned exact bifurcation value (\ref{E2}), the relative error
being $0.057$.

The above consideration addressed a single soliton in the nonlinear coupler.
A cruder version of the VA was used to analyze two-soliton interactions in
the same system \cite{Doty1995}. Accurate numerical results for the
interactions were reported in \cite{PMC1998}. Furthermore, it was also
demonstrated that chains of stable solitons with alternating signs in the
dual-core fiber support the propagation of \emph{supersolitons}, i.e.,
localized collective excitations in the chain of solitons \cite{LuLi}. The
underlying chain may be built of symmetric solitons, as well as of
asymmetric ones with \emph{alternating polarities}, i.e., intermittent
placements of larger and smaller components in the two cores.

\section{Solitons in dual-core fibers: experimental results and the
theoretical framework}

As said above, the experimental observation of solitons in nonlinear optical
couplers was reported much later than such solitons were predicted
theoretically. The spontaneous formation of asymmetric temporal solitons in
the symmetric dual-core fiber was first experimentally demonstrated using a
sample of length $4.3$ cm, with the anomalous-GVD coefficient in each core $%
\beta _{2}=$ $-0.0773$ ps$^{2}$/m in physical units \cite{Ignac1}. The
sample was fabricated in a drawing tower, using an optimally designed
preform. The coupler's switching length, i.e., a characteristic propagation
length necessary for the transfer of an optical signal between the cores, in
the linear regime (this propagation distance is defined as $1/K$, in terms
of Eqs. (\ref{ucoupler},\ref{vcoupler})), was $1.3$ cm.

In the experiment, soliton-like pulses of temporal length $75$ fs, carried
by the standard telecommunication wavelength, $1560$ nm, were coupled into
one core of the double fiber at repetition rate $100$ MHz, while the other
core was left initially empty. In terms of Eqs. (\ref{ucoupler}) and (\ref%
{vcoupler}), this input is modeled by the initial conditions%
\begin{equation}
u\left( \tau ,z=0\right) =a~\mathrm{sech}\left( \eta \tau \right) ,u\left(
\tau ,z=0\right) =0,~  \label{input}
\end{equation}%
with amplitude $a$ and temporal width $1/\eta $. The energy of the input
pulses was varied, in physical units, in the range of $50-250$ pJ. The
results were observed in the form of the transverse intensity distribution
in the output, which was recorded as a camera image in the fiber facet
(including both cores).

\begin{figure}[tbp]
\includegraphics[width=5.5in]{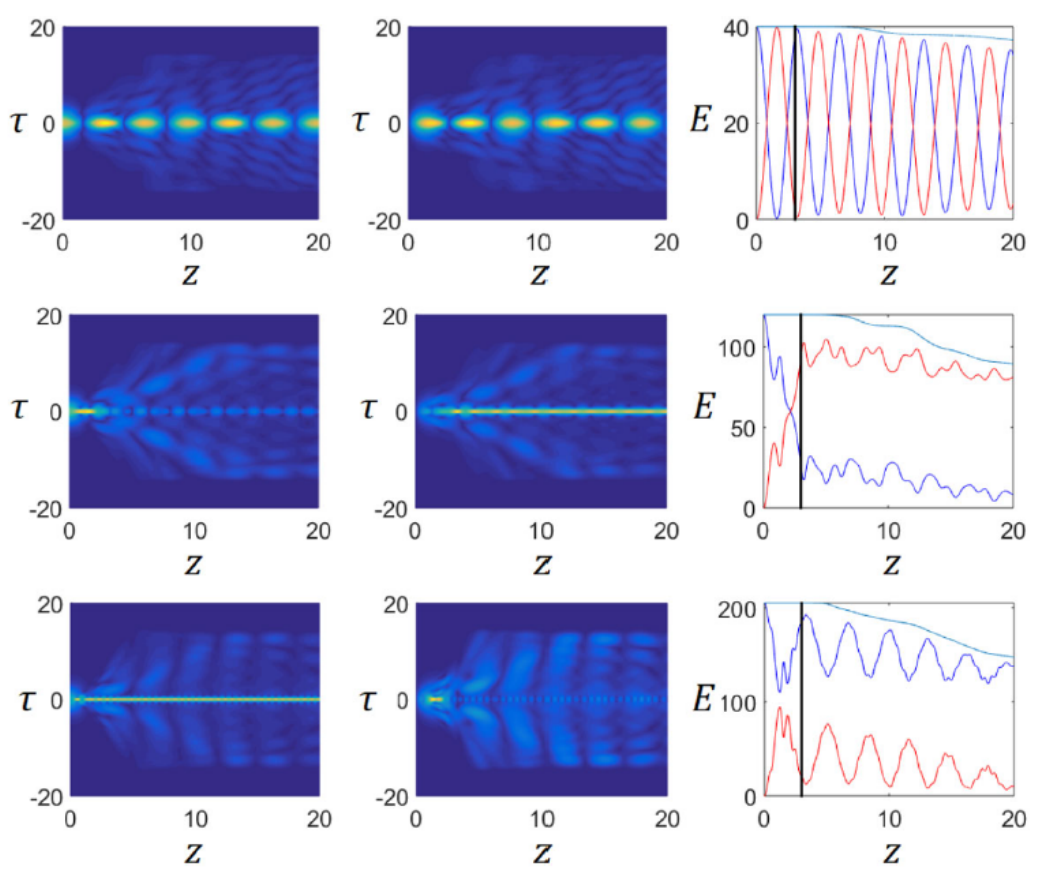}
\caption{Switching between the coupler's cores, produced by simulations of
Eqs. (\protect\ref{ucoupler}) and (\protect\ref{vcoupler}) with initial
conditions (\protect\ref{input}), for three values of the input amplitude, $%
a=1.15$, $2.0$, $2.6$ (from top to bottom), and a fixed inverse width, $%
\protect\eta =0.78$. The left and central columns display spatiotemporal
patterns of the intensities, $\left\vert u(z,\protect\tau )\right\vert ^{2}$
and $\left\vert v\left( z,\protect\tau \right) \right\vert ^{2}$, in the
straight and cross cores, respectively. The blue and red curves in the right
column show the energy in each channel (and the total energy, shown by the
cyan curve) vs. the propagation distance. The top, central, and bottom
panels represent, severally, regimes with periodic inter-core oscillations,
switching into\ the cross core followed by the self-trapping of the
asymmetric soliton in it, and the immediate self-trapping of the asymmetric
soliton in the straight core, respectively. The vertical line at $z=3$
denotes the fiber length in the experiment. The picture, which is borrowed
from the experimental work \protect\cite{Ignac1}, is plotted in scaled
units. }
\label{figOL1}
\end{figure}

\begin{figure}[tbp]
\includegraphics[width=5.5in]{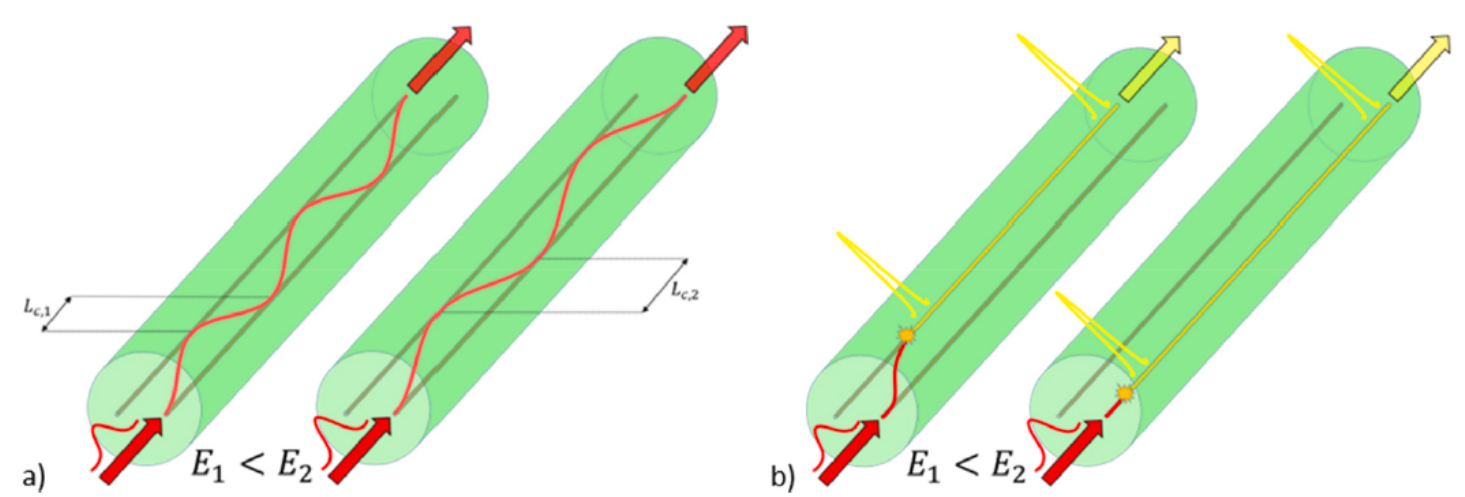}
\caption{A schematic picture of the propagation of an optical pulse
initially coupled into one core of the dual-core coupler, designated by the
pair if diagonal black lines in each drawing (the input pulse is symbolized
by the red arrow at the left edge of each coupler). a) In the regime of
relatively weak nonlinearity, the pulse is periodically hopping between the
cores; larger energy (stronger nonlinearity), $E_{2}>E_{1}$, implies a
longer hopping period. b) In the regime of the relatively strong
nonlinearity, the input pulse self-traps into an asymmetric optical soliton,
which is chiefly trapped in one core, as symbolically shown by yellow
profiles. In the case of smaller energy, $E_{1}<E_{2}$, the input pulse
originally hops from the straight (input) core into the cross one, where it
stays, being compressed into the soliton. In the case of larger energy, $%
E_{2}>E_{1}$, the strongest nonlinearity immediately compresses the input
pulse into the soliton, keeping it in the straight core. The figure is
borrowed from Ref. \protect\cite{Longo}.}
\label{scheme}
\end{figure}

Before displaying the experimental results, it is relevant to produce the
theoretical prediction, provided by simulations of Eqs. (\ref{ucoupler}) and
(\ref{vcoupler}) with initial conditions (\ref{input}). At relatively low
values of the energy, which is $E=2a^{2}\eta ^{-1}$ in terms of input (\ref%
{input}), when the stationary solution gives rise to the stable symmetric
soliton (see Fig. \ref{fig1})), the full numerical simulations produce a
soliton-like pulse periodically oscillating between the two cores, as shown
in the top-row panels in Fig. \ref{figOL1}, and is schematically illustrated
by two plots in Fig. \ref{scheme} a). Similar experimental results for
periodic switching of temporal optical pulses between the cores were
reported in work \cite{Ignac}, for the carrier wavelength $1700$ nm and a
higher energy scale. The increase of the energy leads to a transition from
the oscillatory regime to a quasi-stationary state, in which the soliton has
once switched from the input (\textit{straight}) core into the second (%
\textit{cross}) one, where it self-traps into an asymmetric soliton, with a
smaller residual component remaining in the straight core. This outcome is
demonstrated by accurate simulations in the central row of panels in Fig. %
\ref{figOL1}, and is schematically illustrated by the left plot in Fig. \ref%
{scheme} b). Finally, at the largest considered values of energy, the strong
nonlinearity keeps the input pulse in the straight core, where it directly
self-traps into the corresponding strongly asymmetric soliton. The latter
outcome of the evolution of the input pulse is shown by accurate simulations
in the bottom row of panels in Fig. \ref{figOL1}, and is schematically
illustrated by the right plot in Fig. \ref{scheme} b).

\begin{figure}[tbp]
\includegraphics[width=5.5in]{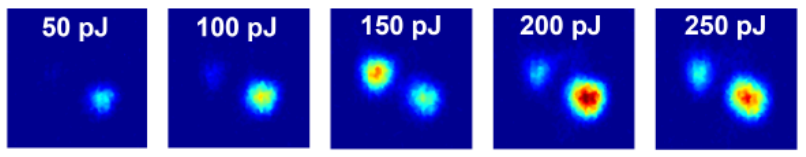}
\caption{A sequence of camera images of the output intensity patterns, taken
in the fiber facet, for different energies of the input and its fixed
temporal width, $75$ fs.The plot is borrowed from work \protect\cite{Ignac1}%
. }
\label{figOL3}
\end{figure}

The theoretically predicted transitions between the different propagation
outcomes which are displayed in Fig. \ref{figOL1}, including the
self-trapping of the asymmetric solitons, were directly observed in the
experiments reported in work \cite{Ignac1}, as shown in Fig. \ref{figOL3}.
Systematic analysis of the experimental data produces a general conclusion:
for the above-mentioned fixed value of the temporal width of input (\ref%
{input}), $75$ fs (in physical units), the gradual increase of the input's
energy $E$ yields the periodic inter-core oscillations, self-trapping of the
asymmetric soliton in the cross core, and, eventually, self-trapping of the
strongly asymmetric soliton in the straight core, in the energy intervals of
$\ 50-100$ pJ, $100-150$ pJ, and $150-250$ pJ, respectively (in physical
units too). In particular, the output patterns presented in Fig. \ref{figOL3}
for $E=50$ pJ and $100$ pJ (in the regime of relatively weak nonlinearity)
correspond to the oscillatory regime. In this case, the energy distribution
between the cores in the output returns, approximately, to the input state
(cf. the top row in the theoretically produced plot displayed in Fig. \ref%
{figOL1}). In the regime of intermediately strong nonlinearity, the
experimentally observed output, shown in Fig. \ref{figOL3} for $E=150$ pJ,
corresponds to the transition from the periodic oscillations to the
self-trapping of the asymmetric soliton in the cross core with residual
oscillations (the length of the fiber is not sufficient to achieve the fully
self-trapped state in this case, similar to what is observed in the
simulations which are displayed in the mid row of Fig. \ref{figOL1}).
Finally, in the strongly nonlinear regime, the experimentally observed
outputs for $E=200$ pJ and $250$ pJ, which are also displayed in Fig. \ref%
{figOL3}, represent the strongly asymmetric solitons with residual
oscillations, which have self-trapped directly in the straight core, cf. the
bottom row in the theoretical simulations displayed in Fig. \ref{figOL1}.
Similar\ experimental results were also observed at other values of
parameters, such as the temporal width of the input.

Some differences between the experimental findings and predictions produced
by the simulations of Eqs. (\ref{ucoupler}) and (\ref{vcoupler}) are
explained by effects of the third-order dispersion and the Raman
self-frequency shift for the optical pulses, which are not included in the
theoretical model.

The findings summarized here and shown in Fig. \ref{figOL3} provide the
first experimental observation of the asymmetric solitons in the symmetric
nonlinear optical coupler, which was theoretically predicted more than
twenty years earlier \cite{Wabnitz,Laval,Maimistov,UNSW,Skinner, Akhm,Akhm2}%
, see also the review \cite{Old}. The demonstration of theoretically
predicted sophisticated features of the symmetry-breaking phenomenology,
such as weakly subcritical character of the SBB, expected according to Fig. %
\ref{fig2}, makes it necessary to perform the experimental studies with a
very high accuracy. It is also relevant to mention other theoretically
explored situations which call for the continuation of the experimental work
in various directions, such as collisions between bright and dark solitons
in optical couplers \cite{PMC1998}, higher-order bright solitons
(breathers), etc.

Solitons in asymmetric dual-core optical fibers have been experimentally
created as well \cite{Ignac2,Ignac3}. Those solitons have a natural
asymmetric structure,

\section{Conclusion}

Dual-core optical fibers is a research area which gives rise to a great
variety of topics for fundamental theoretical and experimental studies, as
well as a plenty of existing and potentially possible applications in
optics, photonics, and plasmonics. While currently employed devices based on
dual-core optical waveguides operate in the linear regime (couplers,
splitters, etc.), the use of the intrinsic nonlinearity offers many options,
chiefly related to the use of self-trapped robust modes in the form of
solitons. In particular, they may be used as data bits in all-optical
integrated schemes of information processing. Then, the dynamical effects
offered by the dual-core waveguides, such as the self-trapping of solitons
in a particular core and switch between the cores, can be employed for the
design of setups controlling the data flow.

In terms of fundamental studies, solitons in dual-core fibers are the
subject of dominant interest, as they offer a setting for the implementation
of many static and dynamical phenomena based on solitons. Furthermore,
optical solitons in dual-core fibers may be used to emulate soliton
phenomenology in other fields of physics, such as Bose-Einstein condensates.

Theoretical studies of solitons in these systems had begun about four
decades ago \cite%
{switch1,switch2,switch3,fiber-coupler2,Trillo,Wabnitz,Laval,Snyder,
Maimistov,Wabnitz2,Akhm,Akhm2,UNSW}. While the basic part of the theoretical
analysis has been essentially completed, there remain many directions for
the extension of the studies. In particular, a natural generalization of
dual-core fibers is provided by multi-core arrays, which allow the creation
of self-trapped modes that are discrete and continuous ones, along the
directions across and along the array, respectively. These modes include
\emph{semi-discrete solitons}, which may be created in many settings\ \cite%
{Tur1,Tur2,Tur3,Hasegawa,RBlit}. Another generalization implies the
transition from 1D to 2D couplers, represented by dual-core planar optical
waveguides. The consideration of the spatiotemporal propagation in these
system makes it possible to predict the existence of novel species of 2D
stable \textquotedblleft light bullets" (spatiotemporal solitons \cite%
{bullet}), including \emph{spatiotemporal vortices} \cite{Dror}.

From many theoretical predictions, only few ones have been realized
experimentally. A brief survey of the experimental results and an outline of
perspectives for the development of the experimental work in this field,
which was not reviewed previously, is the main subject of the present
article. In particular, the fundamentally important species of the
asymmetric solitons in symmetric dual-core fibers, which were predicted long
ago \cite{Wabnitz}-\cite{Skinner}, have been reported in the experiment much
later \cite{Ignac1}. Another noteworthy experimental result is the creation
of semi-discrete \textquotedblleft light bullets" \cite{Jena1}, including
ones with embedded vorticity \cite{Jena2} (actually, in a transient form),
in three-dimensional arrays of fiber-like waveguides permanently written in
bulk samples of silica. Thus, further development of experimental studies in
this vast area remains a highly relevant objective.

\section*{Acknowledgments}

I thank colleagues with whom I have had a chance to collaborate on the
theoretical and experimental topics considered in this article: L. Albuch,
J. Atai, R. Blit, I. Bugar, R. Buzcynski, K. S. Chiang, N. Dror, N. V. Hung,
Y. S. Kivshar, F. Lederer, L. Li, P. Li, M. Longobucco, W. C. K. Mak, V. H.
Nguyen, G. D. Peng, H. Sakaguchi, I. M. Skinner, L. X. T. Tai, R. Tasgal, M.
Trippenbach, I. M. Uzunov, and F. Ye.

\end{document}